\documentstyle[sprocl]{article}

\begin{document}

\title{SPIN-FLAVOUR CONVERSIONS OF NEUTRINOS IN COLLAPSING STARS}

\author{ATHAR HUSAIN}

\address{Departamento de F\' \i sica de Part\' \i culas, Universidade de 
 Santiago de Compostela, E-15706 Santiago de
 Compostela, Espa\~na\\E-mail: athar.husain@cern.ch} 

\maketitle\abstracts{ 
I discuss in some detail the spin-flavour conversions of neutrinos in 
 the  almost isotopically neutral region of collapsing
 stars
alongwith the resulting experimental signatures. In particular, I show that for
realistic magnetic field strength, the observable effects may exist for 
 neutrino magnetic moment, 
$\mu \, \sim \, (2-3)\times 10^{-14}\mu_{B}$ ($\mu_{B}$ is the Bohr magneton)
with relevant neutrino mass squared difference, 
 $\Delta m^{2}\, \sim \, (10^{-9}-10^{-8})$ eV$^{2}$.}

\section{Introduction}

In this contribution,  I  consider a possibility to study  effects of
the neutrino magnetic moments, $\mu $, using  the neutrino bursts \cite{aps}. 
 The spin-flip effects are  determined by the product, $\mu B$, where $B$ is 
 the
strength of the magnetic field. In vacuum (no matter), the spin-flip 
probability 
 may be of the order $1$, if
\begin{equation}
 \mu \int {\rm d}r' B(r')\, \geq \, 1,
\end{equation}
where the strength of the magnetic field
 is integrated along the neutrino trajectory.
Let us suggest the following magnetic field profile
\begin{equation}
 B \simeq B_{0}\left(\frac{r_{0}}{r}\right)^{k},
\end{equation}
with $B_{0} \sim  (10^{12}-10^{14})$ Gauss, $k = 2-3$,
and $r_{0}\sim 10$ km,
where $r$ is the distance from the center of star.
 For sufficiently large upper limit of integration, 
one gets from  (1),
\begin{equation}
\mu B_0 \, \geq \, \frac{k-1}{r_0},
\end{equation}
resulting in  $\mu \, \geq \,   (10^{-15}-10^{-17})\,  \mu_{B}$.
These numbers look encouraging, being  3 - 5 orders
of magnitude below the present upper bound \cite{mom}.
 The presence of dense matter strongly reduces
a sensitivity to $\mu$. Forward neutrino scattering  results
in neutrino level splitting: $\Delta H =  V_{SF}$, where $V_{SF}$ is
the difference of the effective potentials acquired by the left and right
handed neutrino components in matter. Typically $V_{SF} \sim V_0$, where
\begin{equation}
 V_0  \equiv
\sqrt{2}G_{F}n,
\end{equation}
is the total potential, $G_{F}$ is the Fermi
constant and $n$ is the nucleon number density.
The level splitting suppresses the
depth of precession:
 $A_{P} =  \frac{(2\mu B)^2}
{(2\mu B)^2 +  (\Delta H)^2}, $
and to have   $A_{P}\, \geq \, \frac{1}{2}$, one needs, $\mu B  \geq 
 \frac{1}{2} V_{SF}$.
In dense medium this restriction  is much stronger than (3).
The  suppression of the matter density 
 results in the increase of the sensitivity to $\mu$. This is
realized
in the case of the resonant spin-flavour conversion. 
Now the expression for $A_{P}$ in usual notations become
\begin{equation}
 A_{P} = \frac{(2\mu B)^{2}}{(2\mu B)^{2}+\left(V_{SF} -\frac{
 \Delta m^{2}}{2E}\right)^{2}},
\end{equation}
and in the resonance ($V_{SF}=\Delta m^{2}/2E$) one gets $A_{P}\,=\,1$.
This contribution is organized as follows.
In Sect.~2, I describe the effective potential in collapsing stars.
In Sect.~3, the level crossing and the dynamics
of neutrino propagation is briefly considered. 
In Sect.~4,  I discuss  some possible implications of the results.

\section{Isotopically neutral region}

For  spin-flavour conversions,
e.g., $\nu_{eL}\rightarrow \bar{\nu}_{\mu R},\,\bar{\nu}_{eR}\rightarrow
\nu_{\mu L}$, the matter effect
is described by the potential 
\begin{equation}
V_{SF} =   V_0~ (2Y_{e}-1),
\end{equation}
where $Y_{e}$ is the number of electrons per nucleon. 
In isotopically neutral
medium (No. of protons = No. of neutrons),  one has  $Y_{e}\,=\,\frac{1}{2}$,
and according to (6)  $V_{SF} =  0$.
Consequently, the matter effect is determined by the deviation
from the isotopical neutrality.

Below the hydrogen envelope of a  type II supernova  the layers  follow
which consist mainly of
the isotopically neutral nuclei:
${}^{4}$He,  ${}^{12}$C, ... .
Thus the  region  between the hydrogen
envelope and the core  
is almost isotopically neutral. The deviation
from the neutrality is stipulated by  small
abundances, $\xi_{i}$,  of the elements with small
excess of neutrons,
$i = {}^{22}$Ne, $^{23}$Na, ... .
It can be written as
\begin{equation}
(1-2Y_{e}) \approx  \sum_{i}\xi_{i}
\left(1-\frac{2Z_{i}}{A_{i}}\right),
\end{equation}
where $Z_{i}$ and $A_{i}$ are the electric charge and the atomic
number of nuclei ``$i$", correspondingly.
Typically, one gets $\sum_{i} \xi_{i} \leq  10^{-2}$
and $(1-\frac{2 Z}{A}) \leq  10^{-1}$, and
therefore, according to (7): $(1-2Y_{e})\,\leq \, 10^{-3}$.
The value of $(2Y_{e}-1)$ equals
\begin{equation}
 (2Y_e - 1) = \left\{ \begin{array}{ll}
                        0.6-0.7 & \mbox{hydrogen envelope},  \\
                        -(10^{-4}-10^{-3}) & \mbox{isotopically neutral
                                           region},  \\
                        -(0.2-0.4) & \mbox{central regions of star.}
\end{array}
              \right.
\end{equation}
Consequently, in the isotopically neutral region
the potential  $V_{SF}$  is suppressed by more
than 3 orders of magnitude w.r.t  the total potential
$V_0$. The potential is calculated  using the
model of the progenitor with mass  $15 M_{\odot}$
(where $M_{\odot}$ is the solar mass) \cite{ST}.

\section{The dynamics of the neutrino conversions}

Two conditions are essential in studying the dynamics of neutrino 
conversions: level crossing and the adiabaticity.

Consider a system of two massive neutrinos with
transition magnetic moments and vanishing vacuum mixing. 
The diagonal elements of the effective Hamiltonian,
$H_{\alpha}\, (\alpha  =  \nu_{e},  \bar{\nu}_{e},
\nu_{\mu}, \bar{\nu}_{\mu})$, can be written as
\begin{eqnarray}
 H_{e} \equiv 0, &
 H_{\bar{e}} \equiv  -V_0~ (3Y_{e}-1), \nonumber \\
 H_{\mu} \equiv 
 \frac{\bigtriangleup m^{2}}{4E}  - V_0~ Y_{e}, &
 H_{\bar{\mu}}  \equiv
 \displaystyle{\frac{\bigtriangleup m^{2}}{4E}}
 - V_0~ (2Y_{e}-1).
\end{eqnarray}
The {\em level crossing} conditions read as
$H_{\alpha} = H_{\beta}$ ($\alpha \neq \beta$).
 Let us  mention here that the typical relevant neutrino  energy span is
$E\, \sim \, (5-50)$ MeV. 

The {\em adiabaticity condition} in the resonance for spin-flip 
effects reads 
\begin{equation}
 \kappa \equiv  \frac{2 (2\mu B)^{2}}{\pi |\dot{V}_{SF}|}
 \, \geq \,   1.
\end{equation}
Here $\dot{V}_{SF} \equiv  {\rm d}V_{SF}/{\rm d}r$ and  
$\kappa$ is called the resonance adiabaticity parameter.

Let us
define the {\it precession bound} for the product $(\mu B)_{P}$ as
\begin{equation}
 (\mu B)_P = \frac{1}{2} V_{SF}.
\end{equation}
At $(\mu B) = (\mu B)_{P}$, the precession
may  have the depth $A_{P} = 1/2$.
Let us also introduce the
{\it adiabaticity bound} $(\mu B)_{A}$
using the adiabaticity condition (10):
\begin{equation}
 (\mu B)_{A} = \sqrt{\frac{\pi |\dot{V}_{SF}|}{8}}.
\end{equation}
The suppression of  $V_{SF}$
makes the precession effect more profound.
Clearly, $(\mu B)_{P} < (\mu B)_{A}$ in the
hydrogen envelope and $(\mu B)_{P} > (\mu B)_{A}$ in the center
of star. In the isotopically neutral region
one has $(\mu B)_P \sim  (\mu B)_{A}$. 
Instead of $\mu B$, in further discussion I will give  the
estimates of  the magnetic fields
at $\mu = 10^{-12}\mu_B$.

The adiabaticity and the precession  bounds  determined by the density
distribution should be compared with the magnetic field profile
of  star. The strength of  field  profile (2) with $k = 2$ and 
 $B_{0} = 1.5\cdot 10^{13}$ Gauss is practically everywhere
below both the adiabatic, $B_A$, and the precession, $B_P$, bounds
($\mu = 10^{-12}\mu_B$) which means
that  spin-flip effects are rather weak. 

\section{Implications}

It is convenient to introduce  three types of original spectra:
soft, $F_s(E)$, middle, $F_m(E)$, and hard,
$F_h(E)$, which coincide with original $\nu_e$,
$\bar{\nu}_e$,
and $\nu_{\mu}$ spectra, respectively. If there is no neutrino conversions,
then at the exit:
\begin{equation}
F(\nu_e) = F_s,~~~ F(\bar{\nu}_e) = F_m,~~~
F(\nu_{ne}) = 4 F_h.
\end{equation}
Here $\nu_{ne}$ mean $\nu_{\mu}$, $\bar{\nu}_{\mu}$, 
 $\nu_{\tau}$ and $\bar{\nu}_{\tau}$.
Let us find the signatures of the spin-flavour conversions,
as well as the sensitivity of the  $\nu$-burst studies to the neutrino
magnetic moments.

\subsection{Sensitivity to magnetic moments and magnetic fields}

As pointed out in Sect. 1, 
 the suppression of the effective potential (by $3-4$ orders of magnitude)
diminishes the strength of the magnetic field (magnetic moment)
needed for a strong spin-flip effect. The adiabatic and the precession
bounds decrease by $1.5-2$ and $3-4$ orders of magnitude respectively.
I will define
the sensitivity limit $B_s(r)$ in a given point as the
strength of the magnetic field,
for  which   $P_{LZ}(r) \sim 0.75$, where $P_{LZ}$ is the Landau Zener 
probability. 
 According to the estimates in Sect. 3,
the sensitivity bound can be
about 3 times  smaller than  the adiabaticity bound:
$B_s \approx B_A/3$.

For the oscillation solution of the
atmospheric neutrino problem ($\Delta m^2 \sim  10^{-3}$ eV$^2$)
\cite{atm} the resonance is at $r \sim (2 - 3)\cdot10^{-3}~R_{\odot}$,
and $B_s \sim 10^9$ Gauss.  
 At $r \approx  0.3 R_{\odot}$ one gets
$B_s \sim (2 - 3) \cdot 10^4$ Gauss.  This region corresponds to values
 $\Delta m^2 = (10^{-9} -  10^{-8}$) eV$^2$ which
are interesting from the point of view of the resonant spin-flip in the
convection zone of the Sun \cite{mfs}.

\subsection{Bounds on $ \mu B$ from SN1987A}

The spin-flip $\nu_{\mu}\rightarrow \bar{\nu}_{e}$
leads to an appearance of the high energy tail
in the $\bar{\nu}_{e}$ spectrum at the Earth:
\begin{equation}
F(\bar{\nu}_e) = (1-P_s) F_m + P_s F_h~,
\end{equation}
where $P_s$ is the conversion probability.
Note that in (14)
$P_s$ can be close to 1, in contrast with
averaged vacuum oscillation effect which gives $P \leq 1/2$.

The absence of the distortion of the $\bar{\nu}_e$ energy
spectrum, and
in particular,  the absence of the high energy tail,
gives the bound on $\mu B$ as function of $\Delta m^2$.
For a class of supernovae models the  data from SN1987A allow to get
the restriction   $P_s < 0.35$
with the assumption that $P_s$ does not depend on energy
\cite{SSB}. 
The sensitivity limit obtained in Sect. 4.1
gives an
estimate of the upper bound on $\mu B(r)$. 

\section{Conclusion}

For neutrino mass squared differences
$\Delta m^2 < 10$ eV$^2$ (which are 
 interesting for the cosmology, as well as for
 the physics of solar and atmospheric neutrinos) 
 the resonant spin-flavour conversions
($\nu_{e}\rightarrow \bar{\nu}_{\mu}$ etc.)
take place in almost isotopically
neutral region of collapsing star.
In this region  which extends from
$\sim 10^{-3} R_{\odot}$ to
$\sim R_{\odot}$
the deviation from the isotopical neutrality, $2Y_{e}-1$,
can be as small as $10^{-4} - 10^{-3}$.  Correspondingly,
the  matter potential  for the spin-flavour
conversions,  being proportional to ($2Y_{e}-1$), turns out to
be suppressed by 3 - 4 orders of magnitude.
Moreover, the potential changes sign at the inner edge 
of the hydrogen envelope.

The suppression of the
effective potential in the isotopically neutral region
diminishes the  values of ($\mu$B) needed to induce an
appreciable spin-flip effects by 1.5 - 2 orders of magnitude.

\section*{Acknowledgments}
I thank Agencia Espa\~nola de Cooperaci\'on Internacional 
 (AECI), Xunta de Galicia
(XUGA-20602B98) and CICYT (AEN96-1773) for financial support.


\section*{References}
\begin{thebibliography}{99}

\bibitem{aps} For details, see, H. Athar, J. T. Peltoniemi and 
              A. Yu. Smirnov, {\em Phys. Rev.} D {\bf 51}, 6647 (1995); 
              For a latest discussion, see, H. Nunokawa, 
              R. Tom\`as and J. W. F. Valle, astro-ph/9811181 and references 
              cited therein.

\bibitem{mom} G. G. Raffelt, {\em Phys. Rev. Lett.}  {\bf 64}, 2856 (1990).


\bibitem{ST} S. E. Woosley and T. A. Weaver, 
             {\em Annu. Rev. Astron. Astrophys.} {\bf 24}, 205 (1986).

\bibitem{atm} Y. Fukuda {\em et al,} {\em Phys. Rev. Lett.} {\bf 81}, 1562 
              (1998); {\em Phys. Lett.} B {\bf 433}, 9 (1998); 
              {\bf 436}, 33 (1998). 

\bibitem{mfs} See, for an update  discussion, E. Kh. Akhmedov, hep-ph/9705451; 
              M. M. Guzzo and H. Nunokawa, hep-ph/9810408.

\bibitem{SSB} A. Yu. Smirnov, D. N. Spergel, and J. N. Bachall, {\em Phys. Rev.
              D} {\bf 49}, 1389 (1994).

\end{thebibliography}
\end{document}